\documentclass[12pt]{article}
\pdfoutput=1

\usepackage{amsmath,amssymb,graphicx}
\usepackage{cite,slashed}
\usepackage[svgnames]{xcolor}
\usepackage{url}
\def\winv{\Gamma_\text{inv}}

\def\L{\mathcal{L}}
\def\A{\mathcal{A}}
\def\br{\text{Br}}
\def \lsim{\mathrel{\vcenter
     {\hbox{$<$}\nointerlineskip\hbox{$\sim$}}}}
\def \gsim{\mathrel{\vcenter
     {\hbox{$>$}\nointerlineskip\hbox{$\sim$}}}}

\def\bit{\begin{itemize}}
\def\eit{\end{itemize}}
\def\beq{\begin{equation}}
\def\eeq{\end{equation}}

\begin{document}

\begin{flushright}LYCEN 2014-10\end{flushright}\vskip.5cm

\begin{center}
{\huge \bf Heavy Z' : resonant versus non-resonant searches}\vskip.1cm
\end{center}\vskip5mm

\begin{center}
{\bf {Guillaume Drieu La Rochelle}, {Martin Elmer}}
\end{center}\vskip 8pt

\begin{center}
{\it Universit\'e de Lyon, F-69622 Lyon, France; Universit\'e Lyon 1, Villeurbanne;\\
CNRS/IN2P3, UMR5822, Institut de Physique Nucl\'eaire de Lyon\\
F-69622 Villeurbanne Cedex, France} \\
\vspace*{0mm}
{\tt  g.drieu-la-rochelle@ipnl.in2p3.fr, m.elmer@ipnl.in2p3.fr}
\end{center}\vskip5mm

\begin{abstract}
\vskip 3pt
\noindent
{\small Collider searches for new vector-like particles such as $Z'$ have mostly been pursued by looking for a peak in the invariant mass spectrum of the decay products. However off-shell $Z'$ exchange may leave an imprint on other kinematic distributions, leading thus to non-resonant searches. The aim of this paper is to assess, in the context of the LHC, the interplay between resonant ($s$-channel) and non-resonant ($t$-channel) searches for a generic leptophobic $Z'$ model. We show in particular that while non-resonant searches are less sensitive to small couplings, they tend to be more adapted at high masses and large couplings. We discuss our findings both at the level of the current limits and the expectations at higher luminosities.}

\end{abstract}



\section{Introduction}

The past years have certainly done much to establish the Standard Model as a remarkably efficient theory. Among the recent breakthroughs stand the discovery of a light scalar with properties significantly close to the standard Higgs boson\cite{atlas_higgs,cms_higgs}, the absence of new resonances up to the TeV scale\cite{atlas_exo,cms_exo} or precision tests in flavour physics. Nevertheless, they have not put an end to investigations in the realms of New Physics. Indeed there are still many ways leading to New Physics, Supersymmetry, extra dimensions and compositeness encompassing already by themselves hundreds of models, and a large number of dedicated experimental searches. In that respect the first run of the LHC has significantly stirred the field of searches for new particles (see \cite{atlas_exo,cms_exo}), and the second run, with its increase of the center of mass energy, is eagerly awaited in the phenomenology community as a decisive test for many models. However, the amount of analyses done at the LHC is so large that for a given model, the question as to which analyses is the more likely to discover (or exclude) it is a thorny issue.\\

In this paper we have chosen to study the interplay between different types of experimental searches at the LHC in the case of vector like particles. Such new particles typically appear in many extensions of the SM where the gauge group is extended or can emerge as bound states in composite models (as a simple picture, those can be seen as the analog of QCD vector mesons for a new confining force). Another possibility lies in extra-dimensional models, where those spin 1 particles appear as Kaluza-Klein excitations. Reviews of new vector-like particles can be found in \cite{appelquist_02,langacker_0801}. Examples of such models with particular emphasis on LHC phenomenology can be found in \cite{li_0906,salvioni_0909,basso_1002,diener_1006,belyaev_1010,belyaev_1303} for extended gauge groups, \cite{contino_06,barducci_1210,deblas_1211,barducci_1212} for composite models and \cite{agashe_0709,kubik_1209} in extra-dimensional frameworks, among many others. Another widespread possibility, connected to Dark Matter, is the existence of a Dark Matter portal, where the vector boson allows for the communication between SM particles and Dark Matter (see for instance \cite{profumo_1312,tytgat_1401,mambrini_1403}). Although those models cover a large variety of set-ups and motivations, their discovery prospects at the LHC can most often be described in terms of a generic $Z'$ model, which will be our framework for this study.\\

\begin{figure}[ht]
\begin{center}
\begin{tabular}{ccc}
\raisebox{2mm}{\includegraphics{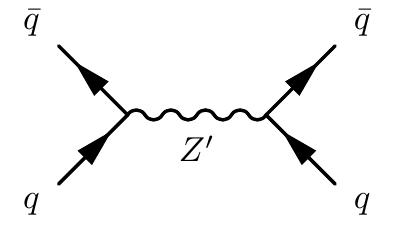}}
  &\hspace*{2cm}&
\includegraphics{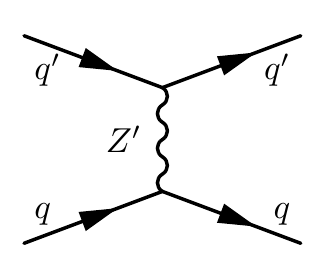}\\
$s$-channel & & $t$-channel
\end{tabular}
\caption{\label{fig:diag}{\small $s$ and $t$ channel diagrams mediating the interactions of a $Z'$ particle with SM quarks. }}
\end{center}
\end{figure}

In most cases it is assumed that the best chance of catching this vector boson at the LHC is through a resonance search, or in other words by producing directly the particle. This is what we refer to as a $s$-channel search, since the main diagram is the $s$-channel one drawn in figure \ref{fig:diag}. In such a case the reach of the LHC will be limited by the probability of the incoming partons to have a center of mass energy higher than the mass of the new particle, which boils down to the pdf (parton distribution functions) of the proton and the beam energy. However, some models can be more elusive to direct production, for instance leptophobic particles escape the cleanest channels at the LHC.\\

When the vector boson escapes direct production, its exchange may still affect the shape of a given observable, without being produced on-shell. An example is a $t$-channel exchange, exemplified in the right diagram of figure \ref{fig:diag}, and for this reason we may refer to those searches as $t$-channel searches\footnote{This dichotomy of $s$ and $t$ searches must be understood as a loose use of the initial word since the diagrams alone are not physically relevant.}. A major point is that processes with many different kinds of initial states can be affected ($uu,ud,\dots$) while direct production requires a particle and its antiparticle as an initial state ($u\bar u,d\bar d,\dots$), leading thus to higher statistics for $t$-channel (non-resonant) searches. The aim of this work is to assess, for a generic vector boson, to which extent the $t$-channel searches can be as powerful, or even better, than $s$-channel searches. Such attempts have already been carried out in the recent years (see \cite{an_1202}), but either with a specific vector boson or with only part of the relevant analyses at the LHC. Here we have opted for a generic discussion, keeping in mind the aim of comparing the sensitivity of different searches on different regions of the parameter space and what would be their future reach. The paper is organized as follows : section 2 introduces the $Z'$ model that we use, section 3 describes the experimental analyses and section 4 discusses our results and leads us to the conclusion. More details on how we recast the analyses to constrain our $Z'$ model can be found in appendix A.


\section{Framework}
We assume that the new vector boson only decays into SM particles or invisible states, in particular we avoid the case of cascade decays through other exotic particles. Its most generic definition is then: it is a spin 1 particle with mass $M$, couplings to all SM particles $g_{XX}$ and width $\Gamma$, which only differs from the partial width to SM particles by an invisible width $\winv$. The free parameters are $M$, $g_{XX}$ and $\Gamma$. Such a generic definition does however involve many final states and as such many analyses from the LHC, we have thus decided to consider only couplings to SM fermions, for simplicity. At this point it is known that if the couplings to leptons are not negligible the constraints from LHC resonant searches in dileptons final states (see \cite{cms_dilepton,atlas_dilepton}) are overwhelming : in such a case it is quite unlikely that a non-resonant search could compete. We focus thus on a leptophobic $Z'$ framework, where the only allowed couplings to SM particles are to quarks. It may seem that the framework has been drastically reduced from its initial scope, however it remains an interesting playground for different models, as shown in \cite{dobrescu_1306}. A last simplification is to assume that the Dirac structure of the couplings is the same as the $Z$ of the Standard Model, leaving free only one scaling parameter, and to assume this scaling factor to be flavour universal. This restriction allows us to use a minimal set of parameters. Our set-up is summarized by the following lagrangian :
\begin{equation}
 \L_\text{N.P.}=\frac{1}{2}F'_{\mu\nu}F'^{\mu\nu}-\frac{1}{2}M^2Z'_\mu Z'^\nu+\kappa\sum_i \left(g^{\text{SM}}_L\bar q_{iL}\slashed{Z}'q_{iL}+g^{\text{SM}}_R\bar q_{iR}\slashed{Z}'q_{iR}\right) + \L_\text{inv.}
\label{eq:lag}
\end{equation}
Here, $\kappa$ refers to the common scaling factor with respect to the Standard Model and $\L_\text{inv.}$ deals with the possible interaction with other particles that would be unseen at the LHC. For all practical purposes we will account for this last term simply via a free parameter $\winv$, so that the total width of the $Z'$ can be larger than what would be expected to SM decay modes, making thus its branching fraction to SM particles lower than 100\%. Our model is thus fully described by 3 parameters: $$\kappa,\ M,\ \winv.$$

Note that the choice of taking the same Dirac structure as for the standard $Z$ is only made for convenience as it lowers the number of free parameters, and we have checked that taking vector-like couplings would not alter our conclusions. We restrict ourselves to $\kappa > 0$ because in this case the $Z'$ interferes constructively with the SM and the expected signals are stronger. Perturbativity imposes an upper bound on $\kappa$, so that the coupling is actually smaller than $4\pi$, which will be the case in our focus range $\kappa<5$.  An important implication of the effective lagrangian postulated in eq. \ref{eq:lag} is that we do not consider mixing with the standard $Z$, and consequently there is no constraint coming from electroweak precision tests. This differs from UV-complete models where all the interactions between both $Z$ are taken into account.\\

Turning now to the LHC analyses, we have considered two representatives of the $s$-channel: the search for resonances in dijets \cite{cms_dijet_resonance} and in $\bar tt$ pairs \cite{Chatrchyan:2012yca}. Each of them consider the case of a $Z'$ with different widths and variable cross-sections, so that we were able to recast their results as a function of our parameters $(M,\kappa,\winv)$, up to adjustments that will be addressed in the following section. Those analyses are the most likely to probe a direct production of a leptophobic $Z'$, and are thus a relevant comparison point for the $t$-channel searches. For the latter we have used the shape analyses of the inclusive jet $p_T$ spectrum \cite{cms_inclusive_pt} and of the dijet angular spectrum \cite{cms_dijet_angular}. In particular, it is interesting to study the interplay between the $p_T$ spectrum analysis and the angular one, which was solely studied in \cite{an_1202}.

The two $t$-channel analyses focus originally on a model of contact interaction between left handed quarks. We validate our recast of those analyses by re-deriving the bounds on this model. The relevant lagrangian for the contact interaction between only left handed quarks of all three families and constructive interference with the SM is   
\beq \label{CI}
\L_\text{C.I.} = -\frac{2 \pi}{\Lambda^2} (\bar{q}_L \gamma^\mu q_L)(\bar{q}_L \gamma_\mu q_L).
\eeq  



\section{Analysis}
We will now describe the four experimental analyses on which we have built our results. Since our set-up (defined in eq.\ref{eq:lag}) is not the one used in those analyses, we had to recast them in a consistent way. In this section we describe only very briefly the different analyses, and postpone the details to appendix \ref{appendixi}.

\subsection{$t$-channel: Dijet angular spectrum}
The dijet angular distribution has been measured by the ATLAS and CMS collaborations to search for contact interactions between quarks \cite{atlas_dijet_angular,cms_dijet_angular}. Both analyses being comparable and giving similar results we considered only one, the CMS angular analysis. CMS uses $2.2 \mbox{ fb}^{-1}$ of $\sqrt{s} = 7 \mbox{ TeV}$ data. The analysis studies the normalized dijet angular distributions for different invariant mass regions $M_{jj}$.
The used angular variable that is $\chi_{dijet} = e^{|y_1 -y_2|}$, where $y_1$ and $y_2$ are the rapidities of the two highest transverse momentum jets. 
The normalized angular spectrum is then given by $\frac{1}{\sigma_{dijet}} \frac{d\sigma_{dijet}}{d\chi_{dijet}}$. 
For the QCD dijet background this normalized angular spectrum is almost flat. New physics like contact interactions or $Z'$ mediated interactions predict a peak of the spectrum at low $\chi$. 

We recast the analysis by calculating both background and $Z'$ signal at leading order and consider only events with large invariant mass $M_{jj}>3\mbox{ TeV}$. Our statistical evaluation follows closely the experimental analysis, as explained in Appendix \ref{appendixi}. We validated our analysis on the same contact interaction model as the CMS study. We find an expected exclusion of $\Lambda^{expct} = 10.2 \mbox{ TeV}$, while the experimental analysis exhibits an expected value of $\Lambda^{expct} = 10.9 \mbox{ TeV}$. We conclude that our analysis is a bit conservative but sufficiently accurate for our goals.


\subsection{$t$-channel: Inclusive jet $p_T$ spectrum}
Our second $t$-channel analysis is the search for contact interactions carried by the CMS collaboration in \cite{cms_inclusive_pt}. The aim is to look for a deviation from the QCD prediction in the inclusive jet $p_T$ spectrum, e.g. events as $p+p \rightarrow j + X$ where $X$ is any collection of particles, using  $5.0 \mbox{ fb}^{-1} $ of $\sqrt{s}= 7 \mbox{ TeV}$ data. The considered events have a partonic center of mass energy $1 \mbox{ TeV} \lsim \sqrt{\hat{s}} \lsim 4 \mbox{ TeV}$. While the QCD background happens to be falling exponentially with increasing $p_T$, the contact interactions will show deviations that are more pronounced at higher $p_T$.

To adapt the analysis to a $Z'$ search we simulate the $Z'$ signal at leading order (as a difference between $Z'$+QCD and QCD only cross-sections) and add it to the QCD background calculated at NLO, taken from the experimental analysis. For the statistical evaluation we use a Bayesian method with a flat prior for the $Z'$ mass to calculate the $CL=95\%$ exclusion mass for a given value of $\kappa$.  We consider systematic errors from renormalisation/factorisation scale and variations of the pdf sets to construct a covariance matrix that takes correlations between bins into account.

We validate our analysis by comparing once again the contact interaction model (\ref{CI}). We compute the expected $95\%$ exclusion expected value to be $\Lambda^{expct} = 15.1 \mbox{ TeV}$ while the CMS analysis obtains a value of $\Lambda^{expct} = 13.6 \mbox{ TeV}$. Thus, our estimation is slightly more affirmative than its true reach, and we will keep this in mind when turning to the $Z'$ case.


\subsection{$s$-channel: Dijet resonance}
For our $Z'$ model the most  obvious decay channel in a resonant $s$-channel search is a light quark-antiquark pair, which would manifest itself as a pair of jets with invariant mass equal to $M$. Such searches have been carried out recently both at ATLAS and CMS, here we will use the analysis described in \cite{atlas_dijet_angular}, which searches the dijet invariant mass distribution for peaks. The analysis sets bounds on the quantity $\sigma\times\br\times\A$ (where $\A$ is the acceptance of the search) on typical $Z'$ particles, the only requirements being that the $Z'$ is produced by $\bar qq$ initial states and that its  shape is approximately Gaussian. Three different resonance widths are considered, namely $\Gamma/ M  = 7\% , 10\% , 15\%$.  This restricts the recast of our analysis to not too wide $Z'$. This is to put into perspective with the $t$-channel case whose reach is independent of the width.\\


\subsection{$s$-channel: $t \bar{t}$ resonance}

Both CMS \cite{Chatrchyan:2012yca,Chatrchyan:2013lca} and ATLAS \cite{Aad:2013nca,TheATLAScollaboration:2013kha} are looking for $Z'$ resonances in the $t \bar{t}$ spectrum. The two top quarks decay into two leptons plus two jets and missing transverse momentum due to neutrinos. We focus on \cite{Chatrchyan:2012yca} as it uses a similar integrated luminosity than the $t$-channel analyses. The analysis constrains any massive neutral vector boson that couples to quarks by putting an upper bound on the product of production cross section times branching ratio to tops  $\sigma \times \br$ as a function of the $Z'$ mass $M$ for two different widths $\Gamma /M= 1.2\% $ and $\Gamma /M =  10\%$. This implies that also this analysis can only be used directly for widths that are not much larger than $10\%$. For larger widths, the exclusion is expected to be weaker, so we have extrapolated an upper bound on the exclusion by taking the limit at $\Gamma /M =  10\%$. For widths in-between $1.2\%$ and $10\%$ we use the less stringent one as a conservative estimate.

\section{Results}

\subsection{Observed results}

\begin{figure}[ht]
\begin{center}
\includegraphics[width=13cm]{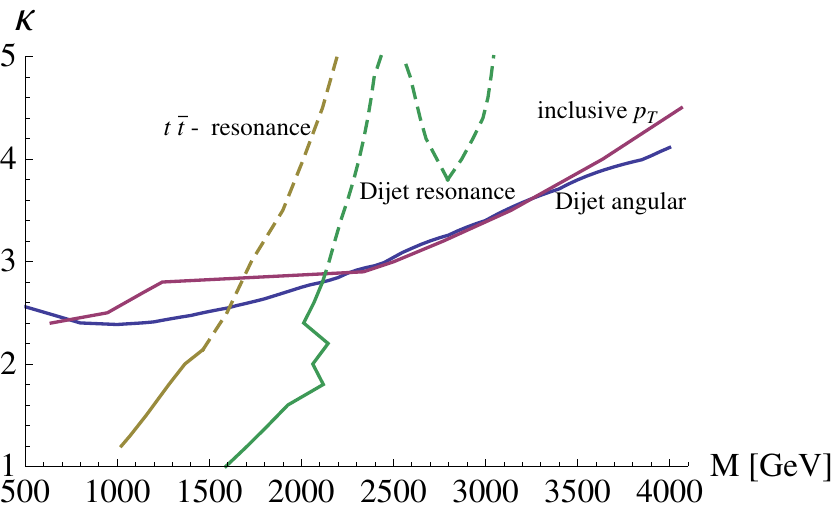}
\end{center}
\caption{\small The observed exclusion at $CL = 0.95\%$ in the plane $M - \kappa$. The angular dijet and $p_t$ spectrum search correspond to the blue and magenta line, respectively.  The brown and green curve are the $t\bar{t}$ and the dijet resonant searches. The dashed lines denote that our extrapolation to large values of $\kappa$ is done in the best case scenario. In reality the exclusions for large $\kappa$ are weaker.  Points on the left upper side of the lines are excluded. \label{fig:obs}}
\end{figure}

We show the observed $CL=95\%$ exclusion lines in the $(M,\kappa)$ plane in figure \ref{fig:obs}. This corresponds to the slice $\winv=0$ of our parameter space, the effect of invisible decay channels being discussed later on. We note that the two $t$-channel analyses give very comparable bounds, which was not granted since they do not yield similar bounds on the contact interaction model. On the $s$-channel side, the dijet search performs better than the $\bar tt$: this is no surprise since it was already the case of a ``standard-like`` sequential $Z'$. In the latter case, part of the exclusion line is plotted as dashed, which corresponds to points where the $Z'$ width is larger than the maximal width constrained by the experimental searches. Broader peaks are harder to distinguish from the QCD background and the real limits are probably weaker than our extrapolation. An important point is that low couplings ($\kappa<2$) are not probed at all by the $t$-channel analyses, whereas $s$-channel can be sensitive if the mass is low enough. Overall, a striking feature is that the shape of $t$ and $s$ exclusions stand significantly apart, revealing a different kind of sensitivity.\\

First, the $t$-channel limits are best understood by looking at the very high and very low mass regimes. In the limit $M\gg \hat{s}$ (where $\hat{s}$ is the partonic energy of the events considered), one notices that the exclusion grows linearly with $M$. This corresponds to the case where the $Z'$ can be integrated out and replaced by a contact interaction, suppressing thus the particular kinematics of a $t$-channel exchange. In this case the contact interaction is parametrized by only one parameter: $\kappa/M$, which justifies the asymptotic behaviour shown in figure \ref{fig:obs}.	On the other hand, for small masses ($M < 1 \mbox{ TeV}$) we observe that the exclusion does not improve with decreasing mass. This effect can be tracked down to the kinematics of a $t$-channel exchange. Unlike the cross section of a contact interaction in the limit $\Lambda \rightarrow 0$,  the signal cross section of a $t$-channel $Z'$ exchange does not diverge for $M \rightarrow 0$ but saturates to a finite value, as the propagator tends towards $\sim 1 /\hat{t}$ ($\hat{t}$ being the Mandelstam variable on partonic level). This is specially relevant for the dijet angular search, since the shape of the distribution becomes flat, just as in QCD, hence the normalized distribution is left unaltered. Since both $t$-channel analyses have a minimal cut on the $\hat{t}$ variable (through $p_T$ in the inclusive $p_T$ analysis and the angle $\chi$ plus the invariant mass $\hat{s}$ in the angular analysis) this also explains that small\footnote{Small must be understood here as electroweak-like. This is in contrast with contact interaction models for instance, where $\kappa$ is much larger than unity.} couplings ($\kappa\sim 1$) are out of reach for the whole mass range, as lowering $M$ under this cut will not increase the signal any more. The first bounds on the $Z'$ mass can be put for $\kappa \ge 2.3$.\\

The inclusive $p_T$ analysis has a non-trivial feature in the intermediate mass region:  the exclusion curve raises suddenly from $1$ to $1.2$ TeV, and then stays horizontal until $2.0 \mbox{ TeV}$. This is due to the fact that, even though the jet spectrum is dominated by processes such as $uu \rightarrow uu$ due to pdf considerations, a small resonance appears in $\bar qq\to\bar qq$ for $\frac14 M \lsim p_T  \lsim \frac12 M$. When this resonance falls into one of the first $p_T$ bins of the analysis, the total shape is significantly altered. This occurs since those bins have the largest number of events, so they weight more in the normalization of the shape. The exclusion then gets stronger, all the more since some of those bins have small uncertainties. When $M$ increases again, the resonance goes to higher bins, its effect gets smaller and the exclusion curve goes back to the usual slope.\\

The $s$-channel analyses obey quite a different behaviour, in particular at high $\kappa$. The sensitivity is dependent on a subtle mix of theoretical and experimental considerations, which are not obviously deduced from figure \ref{fig:obs}. Indeed the theoretical prediction for mass dependence of the cross-section differs significantly from low $\kappa$ to high $\kappa$: in the first case the width stays small hence the integrated cross-section decreases exactly as the pdf does while in the latter the convolution of a large width resonance with the pdf leads to a much smaller decrease. We reach a similar conclusion here as in \cite{diaz_1308}. At first glance this would make the $s$-channels searches quite efficient at high masses, provided $\kappa$ is large. However, experimental analyses are less sensitive to broad resonances, and in particular the model-independent analysis of ATLAS \cite{atlas_dijet_recast} restricts the integration range to $[0.8,1.2]\times M$. This reduces the expected signal and thus limits the reach of the $s$-channel searches. Note that the two experimental $s$-channel searches only constrain particles with a maximal width $ \Gamma/M =10\% , 15\%$ respectively. For larger widths we extrapolate using a best-case scenario  as described in appendix \ref{appendixi} but the real exclusion is very likely to be weaker than our estimate (therefore the dashed line).\\

Comparing $s$-channel to $t$-channel searches one finds that for small $\kappa$ the $s$-channel searches are more efficient. Due to the small width the signal of a $Z'$  in a resonance search is distributed over very few bins which makes it easy to discriminate against the background. For larger $\kappa$ the width increases and the signal gets diluted in a resonant search. The $t$-channel searches are independent of the $Z'$ width and tend to exclude larger masses for larger couplings. In the low mass region $s$-channel searches are dominant because in this region the $t$-channel searches are limited by kinematic effects which do not affect the resonant searches.  However for large masses the resonant searches are limited by the parton distribution functions. To produce an on-shell resonance one needs an antiquark in the initial state which is highly suppressed for large energies, whereas for a $t$-channel exchange the $Z'$ does not have to be on-shell and only valence quarks are needed which are less suppressed by pdf.\\

Extending our conclusion to the whole $(M,\kappa,\winv)$ space is straightforward. $t$-channel analysis are blind to $\winv$, so their exclusions stay identical. $s$-channel analyses will be doubly affected: first because the total width increases and second because the branching ratio to the observable final state decreases. Both effects will contribute to lower the sensitivity of those searches, shifting thus the exclusion lines to lower masses. Note that the second effect occurs only when the width is small as compared to the PDF variation scale, otherwise the narrow-width approximation does not apply and the cross-section starts to be insensitive to the branching ratio. Such an extension does not change the global picture that small values of $\kappa$ are only constraint by resonant searches whereas for $\kappa \gsim 2.5$ non-resonant $t$-channel searches are more efficient. 

\subsection{Projections at higher luminosities}

\begin{figure}[ht]
\begin{center}
\includegraphics[width=13cm]{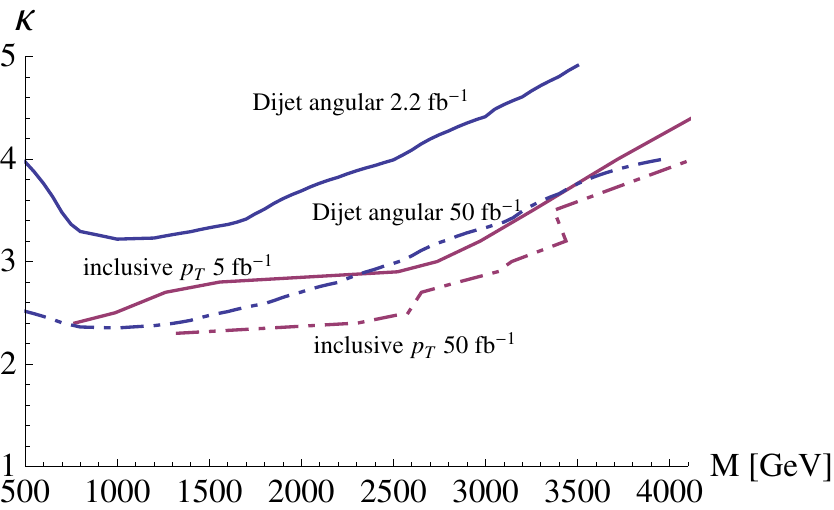}
\end{center}
\caption{\small The expected exclusions at $CL = 0.95\%$ for $2.2 \mbox{fb}^{-1}$ (angular dijet analysis), $5 \mbox{fb}^{-1}$ (dijet analysis) and the expected exclusions for $50 \mbox{fb}^{-1}$ (both analysis) with a constant centre of mass energy $\sqrt{s} = 7 \mbox{ TeV}$. in the plane $M, \kappa$.\label{fig:exp}}
\end{figure}

In figure \ref{fig:exp} we show the expected exclusions for both $t$-channel analyses each once with the luminosity used by the experimental study and once with $50 \mbox{ fb}^{-1}$ at the same centre of mass energy $\sqrt{s} = 7 \mbox{ TeV}$. A compelling feature is that the angular dijet study can still improve the exclusion at higher luminosity since the dominant uncertainty comes from the low statistics for events with $M_{jj} > 3 \mbox{ TeV}$. The inclusive $p_T$ analysis cannot be improved so much by acquiring more statistics since already in its present state the study is dominated by systematic errors. On the $s$-channel side, the increase of luminosity is expected to have a limited impact since the signal is mainly suppressed by pdf.

Comparing figure \ref{fig:obs} and figure \ref{fig:exp} one notices at once that the angular dijet analysis has an observed exclusion that is much more stringent than the expected exclusion. This is consistent with what was observed for contact interactions and is due to the fact that in the two lowest $\chi_{dijet}$ bins  the observed value is below the QCD prediction where new physics gives an enhancement compared to the QCD background.\\

The expectations at 50 fb$^{-1}$ show, albeit in a very crude way, what could be gained by updating the $t$-channel searches with the entire dataset of Run I (counting 25 fb$^{-1}$ for ATLAS and CMS each). Although this is already an important indication to what can be gained by taking into account such analyses, it is also tempting to guess what would come from an increase of the center of mass energy such as will occur for the Run II. The gain may seem obvious for $s$-channels searches since higher invariant masses get accessible, but assessing the ratio of the signal to the background is much more intricate, making a precise prediction tough. For the $t$-channel analyses, the gain depends on the strategy. For the inclusive $p_T$ study, the deviation of the shape to the QCD background increase with $p_T$, so that accessing to higher $p_T$ should enhance the sensitivity. This statement must however be qualified, we have noted that for the data presented in \cite{cms_inclusive_pt}, most of the exclusion was coming from moderate $p_T$ values, due to statistical and systematic uncertainties in higher bins. The angular dijet analysis can be improved by requiring a higher cut on the invariant mass of the dijet. This will reduce the QCD background but keep the signal constant, in the case where $M$ lies above the cut. If $M$ is lower than the cut, there may not be such enhancement.


\section{Conclusion}

New vector particles appear in many a theory beyond the SM and bear a great discovery potential at the LHC. Such states are traditionally searched for as resonances in the invariant mass spectrum of the decay products. However one can also constrain vector particles by their effect on the shape of a given observable without the need of producing the particle on shell. These processes involve most of the time a $t$-channel exchange of the new particle. We compare both methods by testing their exclusion potentials for a sequential $Z'$, with mass $M$, that only couples to SM quarks with couplings proportional to the SM couplings of the $Z$ boson, $\kappa$ being the constant of proportionality.

It turns out that the critical variable that determines the strength of the $t$-channel searches is the coupling strength. Most models proposed in the literature predict $Z'$ couplings to SM quarks of the same order as the SM couplings or even weaker, e.g. in models where the $Z'$ couples to dark matter as well as SM quarks, the couplings to SM must be quite small to avoid exclusion from direct detection experiments. We find that $t$-channel searches can only constrain $Z'$ models when the couplings are sufficiently large $\kappa \gsim 2.3$. For smaller couplings even low masses are not constrained because the propagator then goes like $1/\hat{t}$ and is independent of $M$. In the case of SM-like or smaller couplings, the $s$-channel searches are thus more efficient. The situation changes significantly at large couplings since the resonant searches are hampered by the larger width of the $Z'$, which makes it difficult to distinguish from the background. $t$-channel searches are however independent of the $Z'$ width and tend to exclude larger masses for larger couplings. For couplings $\kappa \gsim 2.5$ we find that $t$-channel searches constrain larger masses than resonant searches.  This effect is also due to the pdf suppression for large masses, which is stronger for $s$-channels than for $t$-channels because antiquarks in the initial state are not needed in the latter one. The possibility of a coupling to an invisible sector will decrease the reach of $s$-channel search, leaving $t$-channel ones unaltered.  \\

\noindent\textbf{Acknowlegments}\\
We would like to thanks Sacha Davidson for stirring our interest in the non-resonant searches and for a careful reading of this draft, H.~Prosper for the details on contact interaction searches and S.~Brochet, V.~Sordini and L.~Sgandurra for useful discussions on the experimental methods.

\appendix

\section{Analysis recasts}
\label{appendixi}
In this section we give more details on the way we recast the experimental analyses in order to constrain our $Z'$ model. For the two $t$-channel studies we compare our recast to the experimental analyses by constraining the same contact interaction models. 

\subsection{$t$-channel: Dijet angular spectrum}
We recast the CMS analysis \cite{cms_dijet_angular} that is looking for new physics effects in the normalized dijet angular spectrum. In the case of pure QCD background the normalized dijet angular spectrum is almost flat whereas new physics like contact interactions of $Z'$ are peaked at low $\chi$. As the most stringent bound comes from dijets with high invariant mass $M_{jj} > 3 \mbox{ TeV}$, we will only focus on these events. The angular distribution is cut into 7 bins in the region $1 \le \chi_{dijet} \le 16$.

 We simulate both our $Z'$ signal and the background at tree level using \texttt{CalcHep} \cite{Belyaev:2012qa} and the \texttt{CTEQ6L} pdf set \cite{Pumplin:2002vw}. The signal is defined by 
\beq
\sigma_{Signal}^{LO} = \sigma_{QCD+Signal}^{LO} - \sigma_{QCD}^{LO}.
\eeq
where the $^{LO}$ superscript stands for Leading Order, indicating that we used tree-level amplitudes. We do not hadronize our final state and assume that every outgoing quark and gluon produces one jet. Our QCD$^{LO}$ angular spectrum is in very good agreement with the QCD$^{NLO}$ spectrum used in the CMS angular analysis. 

We follow the statistical procedure of the experimental analysis by using the same statistical test function for a given distribution $dist$ of the data and a given $Z'$ hypothesis $\Lambda$ consisting of the $Z'$ mass $M$ and the coupling $\kappa$ (the width is irrelevant in this analysis).
\beq
Q(dist,\Lambda) = -2 ln\left( \frac{ L(dist, \Lambda)}{L(dist, 0)} \right)
\eeq
Here $0$ is the background only hypothesis, corresponding to $\kappa=0$. The likelihood $L$ is given by a product of Poisson likelihood functions for each $\chi_{dijet}$ bin, with the total number of events kept fixed. To obtain the p-value of our hypothesis, we perform a large number (500 000) of pseudo-experiments for the $\Lambda$ hypothesis and the $0$ hypothesis in order to compute the quantities $P(Q(dist,\Lambda) \ge Q(data,\Lambda))$  ans $P(Q(dist,0) \ge Q(data,0))$ and finally the $p$-value. The systematic uncertainties are taken from the experimental analysis, up to the correlations between different bins which are not given. We have circumvented the issue by assuming that, since this uncertainty comes mostly from the renormalisation and factorisation scales as indicated by Table 1 of \cite{cms_dijet_angular}, it would be nearly fully correlated, which is supported by the triangular shape of the uncertainty in fig 2 of \cite{cms_dijet_angular}. In any case, the analysis is clearly dominated by statistical errors, hence the implementation of systematic errors does not change much the exclusion limits. A hypothesis $\Lambda$  is excluded at $95\%$ confidence level if $CL_s = p < 0.05$.

We validated our analysis on the same contact interaction model as the CMS study. We find an expected (observed) exclusion of $\Lambda^{expct} = 10.2 \mbox{ TeV}$ ($\Lambda = 13.4 \mbox{ TeV}$), while the experimental analysis exhibits an expected (observed) value of $\Lambda^{expct} = 10.9 \mbox{ TeV}$ ($ \Lambda = 11.7 \mbox{ TeV}$). We conclude that our analysis is sufficiently accurate for our goals.

\subsection{$t$-channel: Inclusive jet $p_T$ spectrum}

The second $t$-channel analysis that we use is the search for a deviation in the one jet inclusive $p_T$ spectrum done by CMS \cite{cms_inclusive_pt}.
The research is restricted to the region where contact interactions have the largest effect, $| \eta | < 0.5$. The events are collected in 20 $p_t$ bins in the region $507 \mbox{ GeV} \le p_T \le 2116 \mbox{ GeV}$.

We take the QCD$^{NLO}$ background prediction from the CMS $p_t$ analysis which was simulated using \texttt{fastNLO} and the \texttt{CTEQ6.6} parton distribution functions \cite{Nadolsky:2008zw}. We define the $Z'$ signal as previously: $\sigma_{Z'}^{LO} = \sigma_{QCD + Z'}^{LO} - \sigma_{QCD}^{LO} $. This signal is added to the QCD$^{NLO}$ background and then compared to the data.

Our statistical procedure follows the Bayesian method with flat prior used by the experimental analysis as a cross-check. We consider only the shape  by normalizing the total number of events to the observed total number of events. We define $\lambda \equiv 1/M^2$ ( $\lambda \equiv 1/\Lambda^2$) for the Z' (contact interaction) model. The CL=95\% exclusion limit for a fixed value of $\kappa$ and for a given observed (or expected) distribution $dist$ of the data is then defined by
\beq
\int_0^{\lambda^{95}} \overline{L_{p_t}} (dist|\lambda, \kappa) d \lambda = 0.95
\eeq 
 where $ \overline{L_{p_t}}$ is the marginal likelihood. The likelihood of a distribution $dist$ characterized by the number of events in every bin $N_j$, with $j \in [1,20]$, given a set of 20 cross section $\vec{\sigma}= (\sigma_j)$, is calculated using the probability density function of a multinomial distribution
\beq 
L_{p_t} (dist, \vec{\sigma}(\lambda, \kappa)) = \frac{N!}{N_1! \cdots N_{20}!} \prod_{j=1}^{20} \left( \frac{\sigma_j(\lambda, \kappa)}{\sigma(\lambda, \kappa)} \right)^{N_j}
\eeq
where $N$ is the total number of events and $\sigma(\lambda, \kappa) = \sum_{j=1}^{20} \sigma_j(\lambda, \kappa)$ is the total predicted cross section. The marginal likelihood is then simply the average over $S = 500 000$ sets of cross-sections sampled from a multivariate Gaussian distribution incorporating systematic uncertainties 
\beq
\overline{L_{p_t}} (dist|\lambda, \kappa) =\frac1S \sum_{a=1}^{S} L_{p_t} (dist, \vec{\sigma}_a(\lambda, \kappa)) \;\;. 
\eeq
We construct a 20 dimensional covariance matrix from the errors on the pdfs and from the normalization and factorization  scale $\mu_{r,f}$. The latter ones are the dominant source of systematic uncertainties  whose effect is to globally shift the predictions in all bins in the same direction. For the variations of $\mu_{r,f}$ we include a $k$-factor to take into account that we calculate the variations a LO and not at NLO for which the dependence on $\mu_{r,f}$ is smaller than for LO. The value of $k = 0.5$ is determined by fitting our systematic errors to the errors of the CMS analysis on the QCD$^{NLO}$ prediction. Using the covariance matrix we sample $S$ sets of errors that we add to the predicted QCD$^{NLO}$ and Signal cross-sections. These modified cross sections are used to calculate the marginal likelihood. 

We validate our analysis by comparing once again the contact interaction model (\ref{CI}). We compute the expected $95\%$ exclusion (observed) value to be $\Lambda^{expct} = 15.1 \mbox{ TeV}$ ($\Lambda = 15.6 \mbox{ TeV} $). The CMS analysis obtains an expected (observed) exclusion of $\Lambda^{expct} = 13.6 \mbox{ TeV}$ ($\Lambda = 14.3 \mbox{ TeV}$). Thus, our estimation is slightly  more affirmative than its true reach, and we will keep this in mind when turning to the $Z'$ case.

\subsection{$s$-channel: Dijet resonance}

The ATLAS analysis \cite{atlas_dijet_angular} gives the excluded $\sigma\times\br\times\A$ as a function of the resonance mass for three different widths, $\Gamma / M = 7\%, 10\%, 15\%$. To set limits on our $Z'$ model we calculate the cross sections for processes of type $p p \rightarrow Z' \rightarrow q \bar{q}$ using \texttt{CalcHep} and the \texttt{CTEQ6L}L pdf set. We apply the same cuts as the experimental analysis and follow their description of how to recast their analysis given in \cite{atlas_dijet_recast} by integrating the cross section only between $0.8 M \le m_{jj} \le 1.2 M$ where $m_{jj}$ is the dijet invariant mass. The intersection of our $Z'$ cross sections with the excluded cross sections from ATLAS gives the $CL=95\%$  exclusion mass.
 To give conservative bounds we compare the obtained cross section to the excluded cross section with the next in size width, e.g. models with $7\% < \Gamma/M \le 10\%$ are constrained using the $10\%$ exclusion values. For models with large widths $\Gamma/M > 15\%$ we can only  compare them to $15\%$ exclusion but keeping in mind that the real exclusions will be weaker. Our extrapolation for large widths corresponds to a best case scenario in which the excluded cross section does not change with increasing widths, which is probably much too optimistic.

\subsection{$s$-channel: $t \bar{t}$ resonance}
To recast the CMS $t \bar{t}$ resonance search \cite{Chatrchyan:2012yca} we compute the cross section of  $pp \rightarrow Z' \rightarrow t \bar{t}$ using \texttt{CalcHep} with the \texttt{CTEQ6L} pdf set. In contrast to the CMS analysis we do not include a $K$-factor for NLO corrections. The way exclusions are extracted is analogue to the dijet resonant analysis.  In our $Z'$ model in the case of $\winv = 0$, fixing $\Gamma=0.012\ M$ yields $\kappa = 0.74$ and $\Gamma = 0.1\ M$ yields $\kappa = 2.1$. Couplings smaller than one are not very interesting for our analysis so we do not consider the exclusion line for $1.2\%$ width. The calculated cross sections are compared to the exclusions at $10\%$. This gives a conservative limit on all widths $\Gamma/M \le 10 \%$. For larger widths the obtained limit is stronger than the real one, corresponding to the best-case scenario.   


\bibliography{biblio}

\end{document}